\documentclass[useAMS,usenatbib]{mn2e}

\def\plotfiddle#1#2#3#4#5#6#7{\centering \leavevmode
\vbox to#2{\rule{0pt}{#2}}
\includegraphics{#1}}
\def\oao{OAO~1657$-$415}
\def\spacing#1{\renewcommand{\baselinestretch}{#1}\large\normalsize}

\newcommand {\Section}[1]{Section~\ref{#1}}

\newcommand {\Figure}[1]{Fig.~\ref{#1}}

\newcommand {\Table}[1]{Table~\ref{#1}}

\newcommand {\Equation}[1]{Equation~\ref{#1}}

\def\captionfigure#1[#2]#3{
 \def\captionlabel{#1}
 \def\captionlistentry{#2}
 \def\captionheading{#3}
 \begin{figure}}

\def\endcaptionfigure{
 \spacing{1}
 \caption [\captionlistentry]{\captionheading}
 \label {\captionlabel}
 \end{figure}}

\title[{\sl ASCA} Observations of OAO 1657$-$415 and its Dust-Scattered X-Ray Halo]{{\sl ASCA} Observations of OAO 1657$-$415 and its Dust-Scattered X-Ray Halo}

\author[M. D. Audley et al.]{Michael. D. Audley$^{1,2}$\thanks{E-mail: audley@mrao.cam.ac.uk}, Fumiaki Nagase$^3$, Kazuhisa Mitsuda$^3$, 
\newauthor
Lorella Angelini$^{4,5}$, \& Richard. L. Kelley$^4$\\
$^1$UK Astronomy Technology Centre, Royal Observatory, Blackford Hill, Edinburgh EH9 3HJ\\
$^2$Present address: Cavendish Laboratory, University of Cambridge, Madingley Road, Cambridge CB3 0HE\\
$^3$Institute of Space and Astronautical Science, 3-1-1 Yoshinodai, Sagamihara, Kanagawa 229-8510, Japan\\
$^4$Code 662, NASA / Goddard Space Flight Center, Greenbelt, MD 20771, USA\\
$^5$Department of Physics and Astronomy, Johns Hopkins University, Baltimore, MD, 21218, USA}

\begin{document}



\maketitle

\label{firstpage}

\begin{abstract}
We report on two {\sl ASCA} observations of the high-mass X-ray binary
pulsar \oao.  A short observation near mid-eclipse caught the source in a low-intensity state, with a weak
continuum and iron emission dominated by the 6.4-keV
fluorescent line.  A later, longer observation found the source in a
high-intensity state and covered the uneclipsed through mid-eclipse
phases.  In the high-intensity state, the non-eclipse spectrum has an absorbed
continuum component due to scattering by material near the pulsar and 80
per cent of the fluorescent iron emission comes from less than
19~lt-sec away from the pulsar.  We find a 
dust-scattered X-ray halo
whose intensity decays through the eclipse.  We use this halo to
estimate the distance to the source as $7.1\pm1.3\rm\ kpc$.  
\end{abstract}


\section{Introduction}
\oao\ is one of the most poorly studied eclipsing high mass 
X-ray binary systems, and yet is potentially one of the most 
interesting. The reason this system is not so well known is because it took
15 years for its nature as a high mass X-ray binary to be revealed. \oao\ was 
discovered with the {\sl Copernicus} satellite \citep{Poli78}.  
A subsequent observation with {\sl HEAO-1} revealed a 
pulse period of 38 s \citep{WP79}. The 3--30~keV {\sl Ginga} 
spectrum \citep{Kama90} is a typical X-ray 
pulsar spectrum with a power law ($\alpha \sim 0.6$), a high energy 
cutoff ($E_c \sim 5$~keV and $E_f \sim 17$~keV) and an iron line at 
6.6~keV (equivalent width $\sim$ 240 eV). 

For many years it was unclear if \oao\ was a low-mass or a 
high-mass system. The mystery was cleared up when the {\sl CGRO} BATSE 
discovered a $\sim 10.4$-d binary orbit based on long term 
monitoring of the pulse period in the 20--40~keV band \citep{Chak93}. 
A $\sim 1.7$ day eclipse of the X-ray source by its 
companion was also seen. This conclusively showed \oao\ to be a classic 
eclipsing massive X-ray binary system, probably similar to Vela X-1, making 
it the seventh eclipsing X-ray pulsar system to be 
found.  \citet{Chak93} used the orbital parameters to infer that the 
companion is a supergiant of spectral class B0--B6. 
From the observed value of the spin period and its derivative during the 
spin-up interval they deduced an X-ray luminosity of $\ga 1.6\times
10^{37}$~ergs~s$^{-1}$, and thus a distance of $\ga
11\rm\ kpc$. 
A more precise X-ray location obtained with Chandra allowed \citet{Chak02} to
make an infrared identification of the companion.  The infrared
properties of the companion are consistent with a highly-reddened B
supergiant at a distance of $6.4\pm1.5\rm\ kpc$, implying an X-ray luminosity
of $3\times10^{36}\rm\ erg s^{-1}$.  

\oao's position on the Corbet diagram \citep{Corb86} suggests that it may not be a typical system.
The pulse periods of 
high-mass X-ray binaries (HMXRB) are distributed between 0.069 and 835~s, with no evidence 
for a clustering at any particular period. For the Be star systems 
there is a strong correlation between orbital period, $P_o$, and spin
period, $P_s$. The supergiant systems have no strong dependence 
of orbital period on spin period. \oao's 38~sec pulsar with its 10.4-d orbital period lies between the 
Be and supergiant systems 
on the Corbet diagram. The period measurements from {\sl RXTE}, BATSE, and 
previous observations give a steady spin-up timescale of 125 yr \citep{Bayk97} with short-term fluctuations \citep{Bayk00}. This system 
may be in a short-lived phase where it is changing from a wind-accreting 
system like Vela X-1 to a disk-accreting system like Cen X-3. 
Thus, \oao\ provides 
an opportunity to test wind and/or disk accretion 
theory on a system which is in a transition between the two states.

\section{The Observations}
\oao\ was observed near the middle of its eclipse on 1994 March
24, yielding 22~ks of data.  This observation covered orbital phases
$-0.001\mbox{--}0.074$. It was observed later (near 1997 September 17)
through its eclipse ingress for 200~ks.  This observation, covering
orbital phase $-0.210\mbox{--}0.011$, yielded 67~ks of SIS data.
The orbital phases in this paper are calculated using the ephemeris of \citet{Bild97}.  For both of the observations the SIS was operated in one-CCD bright mode and the GIS in PH mode.  The eclipse spectra were significantly different between the two observations.  We discuss these differences in \Section{intensitystates}.

\section{Results}
\subsection{Timing analysis}
\subsubsection{The Light Curve}
\begin{figure}
\plotfiddle{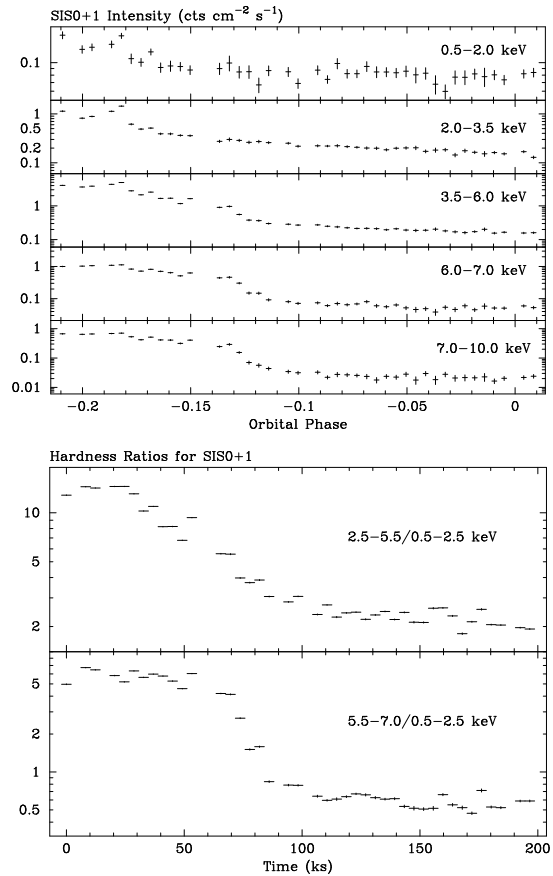}{328.214pt}{0}{66.0}{66.0}{-180.286pt}{-131.418pt}
   \caption
   { \label{lightcurves}    
Top: Energy-resolved SIS lightcurves observed during the second {\sl ASCA} observation.  The slow decay of the 
intensity in the low-energy bands through the eclipse is due to the decaying 
dust halo.  Bottom: SIS hardness ratios. Note that the lightcurves are
plotted against orbital phase, while the hardness ratios are plotted
against time from the start of the observation.  Both are plotted over the same range to allow direct comparison.} 
\end{figure}

Energy-resolved SIS lightcurves and hardness ratios for the second observation are shown in \Figure{lightcurves}.  The eclipse transition is 
extended, with the high-energy band becoming fully obscured around
orbital phase 
$-0.09$.  
The flux in the low energy bands decays gradually in the eclipse, suggesting the presence of a persistent soft excess.

\subsubsection{The pulse profile}
We derived a barycentric pulse period of $37.39\pm0.01\rm\ s$ from the
non-eclipse part of the 1997 observation.
The pulse profile from the 1997 observation is qualitatively similar to the 1--20~keV pulse profile observed by \citet{Kama90} with the GINGA LAC.  
Most of the power is in the first harmonic.  The pulsed fraction also
increases with energy, as would be expected for a source with such a
high absorbing column density ($\sim5\times10^{22}\rm\ cm^{-2}$).  The
dip which occurs after the main peak (labeled `A' in
\Figure{profiles}) seems to be sharper in the present data than in the
GINGA data.  When the 1--10~keV GIS2 and GIS3 data from the first
30~ks of the observation are folded into 256 pulse-phase bins, dip A
has a Gaussian width ($\sigma$) in pulse phase of  $0.0256 \pm 0.0054$
at 90 per cent confidence.  This corresponds to a duration of $0.96 \pm
0.20\rm\ s$. 
The main dip
corresponding to the pulse minimum (labeled `B' in \Figure{profiles}) has a width of $0.097\pm0.019$ in pulse phase,
or a duration of $3.6\pm0.7\rm\ s$.  
Sharp dips in the pulse profile have also been seen in Vela~X-1 \citep[e.g.][]{Choi96}, LMC~X-4 \citep[e.g.][]{Levi91}, and GX~1+4 \citep[e.g.][]{Dota89}.  The pulsed fraction in the pulse profile of \oao\ increases with energy,
suggesting that the pulsating component of the continuum is subjected to more absorption than the non-pulsating component.
  
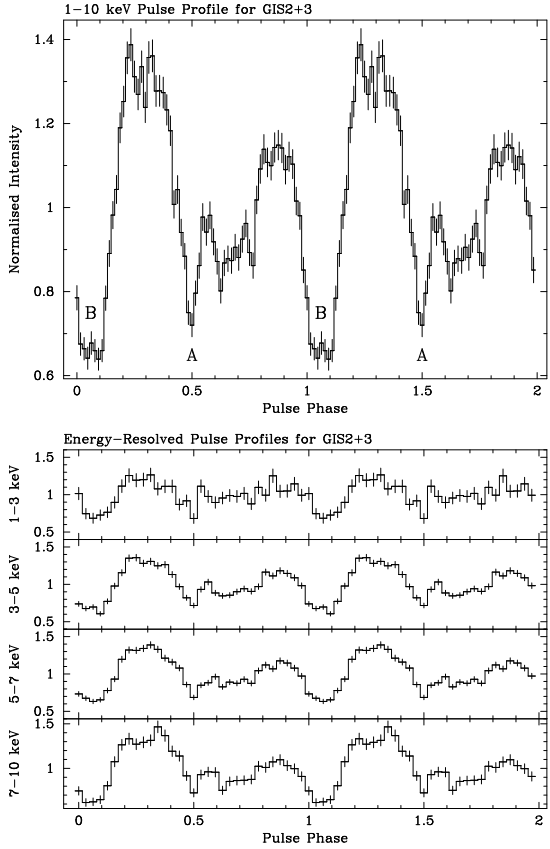
\begin{captionfigure}{profiles}[]{Top: GIS 2+3 energy-averaged pulse
    profile.  The dips in the profile labeled `A' and `B' are
    discussed in the text.  Bottom: GIS 2+3 energy-resolved pulse
    profiles.  
}
\plotfiddle{figure2.epsi}{316.258pt}{0}{63.8}{63.8}{-180.127pt}{-126.886pt}
\end{captionfigure}

\subsection{Spectroscopy}
The standard screening criteria were applied to the data and GIS spectra 
were extracted from circular regions with radii of 6~arcmin.
Because this source is near the Galactic plane, where contamination from 
the Galactic ridge can be significant \citep[e.g.][]{Koya89a}, the standard 
background 
spectra accumulated from blank-sky observations could not be used.
Thus, we extracted background spectra from source-free regions of the detectors.
To account for the azimuthal dependence of the XRT spatial response, 
background spectra for the GIS data were extracted from circular regions 
whose centres were obtained by reflecting the source coordinates through 
the optical axis.  To avoid oversubtraction of the source, only eclipse 
data were used for the background, and the part of the background region 
less than 8~arcmin from the source was excluded.
SIS Bright mode spectra were extracted from 4-arcmin circles around the source.
Background spectra were extracted from the part of the chip more than 
6~arcmin away from the source, using only eclipse data.

\subsubsection{The non-eclipse spectrum}
Referring to \Figure{lightcurves}, the source is essentially uneclipsed for the first 30~ks of the observation, or orbital phases $\phi=-0.210\mbox{--}-0.177$.
We extracted spectra for this interval, added the two SIS and GIS detectors, and fitted a model of the form
\begin{eqnarray}
I(E) &=& e^{-\sigma{E}N_{Hi}}(I_dE^{-\alpha}+e^{-\sigma{E}N_{Hl}}I_sE^{-\alpha} + I_hE^{-(\alpha+2)} + \nonumber\\
&&I_{\rm fl}e^{-(E-E_{\rm fl})^2/2\sigma_{\rm fl}^2} 
+ I_{\rm rc}e^{-(E-E_{\rm rc})^2/2\sigma_{\rm rc}^2})
\label{general}
\end{eqnarray}
simultaneously to the SIS and GIS data.
The continuum is represented by a power law plus a heavily absorbed
power law with the same photon index, $\alpha$.  We interpret these as representing direct 
and scattered emission from the pulsar.  As Thompson scattering is a colourless 
process the  power laws have the same index.  To model the dust-scattered halo we included a power law with a photon index $\alpha+2$
and the same absorption column density as the direct emission power law.  We included two Gaussian emission lines, due to fluorescence of cold iron and recombination of H- and He-like iron, labelled with the subscripts fl and rc, respectively.  

The best-fitting model to the non-eclipse SIS and GIS data is shown in
\Figure{noneclipse}.  The fit is not formally acceptable, with
$P_{\nu}(\chi^2_\nu)=6.9\times10^{-3}$.  Most of the contributions to
$\chi^2$ come from the GIS data.  This model fits the SIS data well,
with $P_{\nu}(\chi^2_\nu)=0.34$.

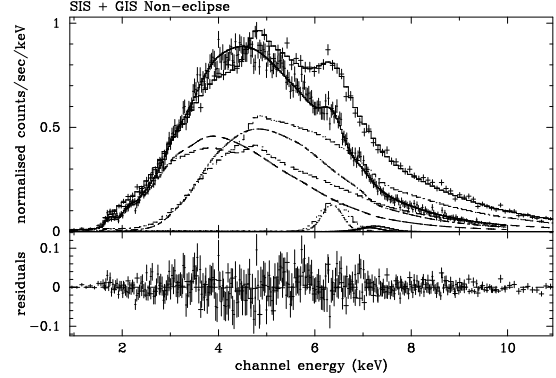
\begin{captionfigure}{noneclipse}[]{Non-eclipse SIS and GIS spectra fitted to the model of Equation~\ref{general}.}
\plotfiddle{figure3.epsi}{140.214pt}{270}{30.1}{30.1}{-108.620pt}{168.798pt}
\end{captionfigure}

\subsubsection{Orbital dependence of the spectrum from the 1997 Observation}

In order to follow the evolution of the spectrum through the eclipse, we divided the data into time intervals based on inspection of \Figure{lightcurves} and attempted to find a consistent model that could be used 
to parameterise the orbital dependence.  
The data were divided into the following five intervals: 
non-eclipse ($\phi=-0.210\mbox{--}-0.177$), pre-eclipse ($\phi=-0.177\mbox{--}-0.144$), ingress ($\phi=-0.144\mbox{--}-0.100$), early-eclipse ($\phi=-0.100\mbox{--}-0.044$), and mid-eclipse ($\phi=-0.044\mbox{--}+0.011$).


\Figure{spectra} shows the SIS spectra at different orbital phases.
We see the iron line equivalent width increase as the eclipse ingress
progresses and the continuum decays, while the spectrum becomes harder due to increased absorption.  However, it is clear that there is a soft component which remains, even as the hard part of the spectrum continues to decrease.  In \Figure{spectra} the 1994 mid-eclipse spectrum is also plotted.  The soft component is clearly much weaker in this observation.
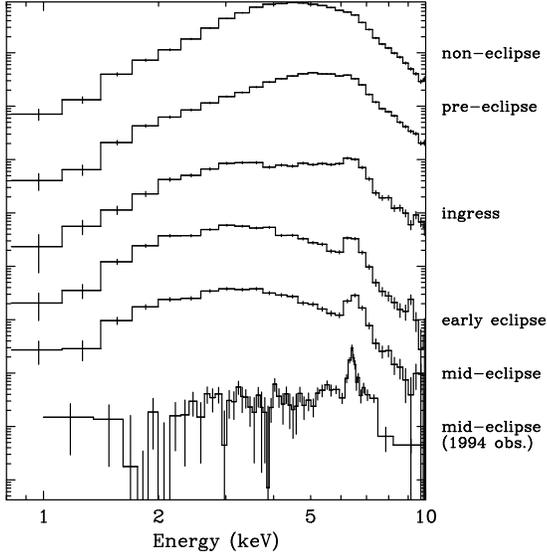
\begin{captionfigure}{spectra}[]{SIS0+1 spectra at different orbital phases.  The orbital phases of the spectra correspond, from top to bottom, to those in \Table{sis01ggis23g_wanponwapopogaga_grosser}.  The mid-eclipse spectrum from the 1994 observation is also plotted at the bottom.  For clarity, the intensity of each successive spectrum has been multiplied by a factor of 0.1.}
\plotfiddle{figure4.epsi}{207.577pt}{270}{41.8}{41.8}{-140.542pt}{240.989pt}
\end{captionfigure}

\Table{sis01ggis23g_wanponwapopogaga_grosser} shows the results of fits to a model of the form shown in \Equation{general}.  In these fits we froze the interstellar column density $N_H$ and the photon index of the direct beam from the pulsar $\alpha$ at the pre-eclipse values.  Otherwise, we were not able to extract meaningful confidence intervals from the fits to the early- and mid-eclipse phases as parameters became degenerate.  As the eclipse progresses, the direct and Thomson-scattered emission components are replaced in the best-fitting model by the halo component.  We thus conclude that the eclipse spectrum is dominated by the dust-scattered halo.

In order to 
investigate the orbital dependence of the iron line complex, we again
split up the data but this time into 19 finer time intervals.  
The intervals were chosen to provide a compromise between counting
statistics and time resolution. The exposures were typically 4--6~ks
in the early parts of the observation, rising to 16~ks for the
mid-eclipse phase.  The lowest number of counts in any spectrum was
425.  The spectra were rebinned so that each bin contained at least 30 counts.
Each spectrum was fitted to a model which had an absorbed power law
continuum and the fitting range was restricted to the 5--8~keV energy
band.  Because of the degeneracy between the line and continuum
parameters, the absorption column density was fixed for the fits.  The power law photon index was allowed to vary.

\Figure{wapoga} shows the results of these fits when the iron line complex was modelled by a single Gaussian.  When the line width was allowed to vary, both the line width and the line energy contained considerable scatter due to off-diagonal elements in the curvature matrix linking them.  Thus, the line width was fixed at 0.25~keV (the observation-average value) for the fits.  The line centroid energy increases from 6.45~keV in the pre-eclipse phase to 6.55~keV at mid-eclipse.  The line intensity drops by an order of magnitude as the pulsar is eclipsed. 
To determine the time taken for this transition to occur, we fitted a
simple continuous model to the intensity which assumes that the
pre-eclipse and eclipse intensities are constant and that the
intensity varies linearly during the ingress.  We find that the
transition occurs in $20\pm9\rm\ ks$, corresponding to a projected
distance of $12\pm6$~lt-sec.

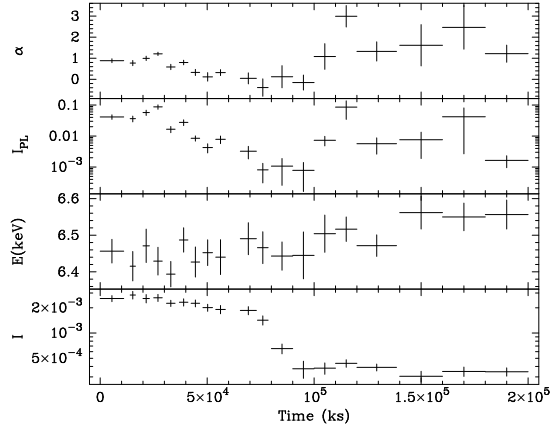
\begin{captionfigure}{wapoga}[]{Orbital dependence of the iron line energy and intensity.  The power law photon index is $\alpha$, $I_{PL}$ is the power law intensity in $\rm cts\ s^{-1}\ cm^{-2}\ \mbox{at}\ 1\ keV$, $E$ is the Gaussian line centroid energy, and $I$ is the line intensity in $\rm cts\ cm^{-2}\ s^{-1}$.}
\plotfiddle{figure5.epsi}{158.876pt}{270}{32.0}{32.0}{-126.044pt}{184.822pt}
\end{captionfigure}


We fitted the iron-line region of the orbitally-resolved spectra with
two narrow lines at 6.4 and 6.97~keV.  
As shown in \Figure{wapogagaga}, the two lines each undergo a partial eclipse.  This means that the regions emitting these lines are larger than the companion star.

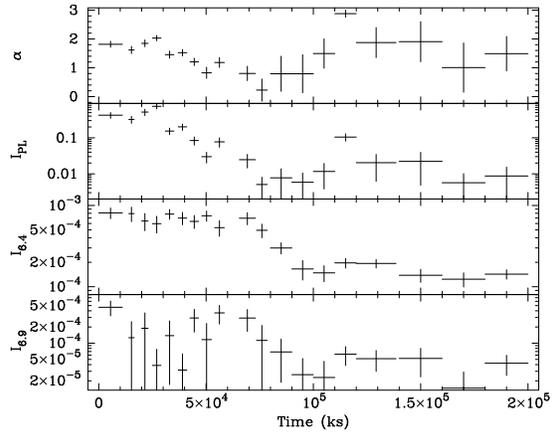
\begin{captionfigure}{wapogagaga}[]{Orbital dependence of the iron
    line intensities for two narrow lines at 6.4 and
    6.97~keV.  The power law photon index is $\alpha$, $I_{PL}$ is the continuum power law intensity in $\rm cts\ cm^{-2}\ s^{-1}\ keV^{-1}$ at 1~keV, and the line intensities are in $\rm cts\ cm^{-2}\ s^{-1}$.}
\plotfiddle{figure6.epsi}{160.664pt}{270}{32.1}{32.1}{-126.764pt}{185.406pt}
\end{captionfigure}

By looking at how long it takes for the individual iron lines to
become eclipsed we can estimate the size of the emission regions for
each of the ionic species (neutral and hydrogen-like iron).  When we fit our simple intensity model to
the line intensities in \Figure{wapogagaga} we find that the line
intensities decrease by factors of $4.5\pm0.6$ and
$2.2\pm1.2$, in intervals of $30\pm19$ and $>23\rm\ ks$,
for the 6.4 and 6.97~keV lines, respectively. These time
intervals correspond to projected distances of $19\pm12$,
 and $>14$~lt-sec, respectively. Thus, the 6.4 keV line
undergoes a deeper and more abrupt eclipse than the 6.97~keV
lines.  80 per cent of the 6.4~keV emission is located within 19~lt-sec of 
the neutron star.  

\begin{table*}
\centering
\begin{minipage}{140mm}
\caption{Best-fitting model parameters for fits to the orbital phase-resolved GIS and SIS data. 
The model is of the form shown in \Equation{general}.  The interstellar absorption column and the photon index of the direct power law are fixed at the pre-eclipse values ($N_H=7.61\pm 0.35\rm\ cm^{-2}$ and $\alpha=1.361\pm0.043$). Note how the direct and scattered continuum components ($I_d$ and $I_s$) are replaced by the dust-scattered halo component ($I_h$) as the eclipse progresses.\label{sis01ggis23g_wanponwapopogaga_grosser}}
\begin{tabular}{@{}lccccc@{}}
\hline
Parameter&non-eclipse&pre-eclipse&ingress&early-eclipse&mid-eclipse\\
\hline
SIS Exposure (ks)&20.8&24.8&16.0&40.5&29.0\\
SIS count rate (cts s$^{-1}$)&$3.30\pm0.01$&$1.62\pm0.01$&$0.451\pm0.005$&$0.205\pm0.002$&$0.142\pm0.002$\\
$I_d$ ($10^{-3}\rm\ cts\ cm^{-2}\ s^{-1}$ at 1 keV)&$33.5 \pm 2.2$&$8.8 \pm 1.4$&$2.90 \pm 0.61$&$0.39 \pm 0.35$&$< 47$\\
$N_{Hl}\rm\ cm^{-2}$&$14.13 \pm 0.71$&$30.0 \pm 1.8$&$61.3 \pm 6.0$&$87 \pm 54$&$35 \pm 12$\\
$I_s$ ($10^{-3}\rm\ cts\ cm^{-2}\ s^{-1}$ at 1 keV)&$92.9 \pm 3.2$&$95.2 \pm 3.3$&$30.9 \pm 3.1$&$2.3\pm 1.3$&$2.07 \pm 0.48$\\
$I_h$ ($10^{-3}\rm\ cts\ cm^{-2}\ s^{-1}$ at 1 keV)&$< 13.1$&$53.3 \pm 1.0$&$52.3 \pm 5.6$&$35.8 \pm 3.5$&$56.0 \pm 2.8$\\
$E_{\rm fl}\rm\ (keV)$&$6.398 \pm 0.021$&$6.356 \pm 0.056$&$6.361 \pm 0.053$&$6.44 \pm 0.47$&$6.366 \pm 0.050$\\
$\sigma_{\rm fl}\rm\ (eV)$&$195 \pm 54$&$227 \pm 88$&$< 139$&$< 131$&$< 149$\\
$I_{\rm fl} (10^{-4}\rm\ cts\rm s^{-1})$&$15.1 \pm 2.3$&$10.1 \pm 5.3$&$2.54 \pm 0.63$&$1.37 \pm 0.48$&$1.94 \pm 0.90$\\
$EW_{\rm fl}$ (eV)&&&&&\\
$E_{\rm rc}\rm\ (keV)$&$7.15 \pm 0.14$&$7.05 \pm 0.45$&$6.638 \pm 0.038$&$6.87 \pm 0.19$&$6.76 \pm 0.16$\\
$\sigma_{\rm rc}\rm\ (eV)$&$< 322$&$< 0.350$&$214 \pm 135$&$233 \pm 116$&$207 \pm 77$\\

$I_{\rm rc} (10^{-4}\rm\ cts\ s^{-1})$&$3.1 \pm 1.8$&$< 8.95$&$3.80 \pm 0.78$&$1.04 \pm 0.52$&$1.07 \pm 0.73$\\

$EW_{\rm rc}$ (eV)&&&&&\\
\hline
$\chi^2_\nu/\mbox{d.o.f}$&1.210/367&1.106/355&1.046/264&1.275/248&0.884/174\\
$P_\nu(\chi^2_\nu)$&$3.52\times10^{-3}$&$0.0826$&0.292&$2.17\times10^{-3}$&0.862\\
\end{tabular}
\end{minipage}
\end{table*}

\subsubsection{The mid-eclipse spectrum}
For this orbital phase we have data from both the 1997 and 1994 observations.  We therefore discuss these together in this section.
For the 1994 observation the intensity in the 2--10~keV band is 
$6.5\times10^{-12}\rm\ erg\ cm^{-2}\ s^{-1}$.  The iron line's equivalent width is 5.68~keV.  
The continuum is weak but a single power-law continuum is not acceptable.  A model which gives an 
acceptable fit has a continuum composed of an absorbed power-law and a
second power-law whose photon index is constrained to be equal to that of the first one plus 2 with less absorption, and two
Gaussian emission lines whose centroid energy and width are free 
(see \Figure{my_bins_oneline_1.4_plot}).  Note that this model has a
simpler continuum than the model described by \Equation{general} and may be expressed as
\begin{eqnarray}
I(E) &= e^{-\sigma{E}N_{Hi}}(I_hE^{-(\alpha+2)}+
e^{-\sigma{E}N_{Hl}}I_sE^{-\alpha}\nonumber\\
& + \sum_{i=1}^NI_ie^{-(E-E_i)^2/2\sigma_i^2})
\label{my_bins_oneline_1.4_equation}
\end{eqnarray}
for the case where there are $N$ Gaussian emission lines.  Here
$N_{Hi}$ is the interstellar column density and $I_h$ is the intensity
of the power law due to the dust halo.  $N_{Hl}$ is the local column
density for the absorbed power law, due to scattering of the direct
pulsar emission in the stellar wind.  The absorbed power law's intensity is $I_s$.  The photon
index of the absorbed, harder power law, $\alpha$, was fixed at 1.4 as it was found to be unconstrained by the data.  The best-fitting parameters are shown in \Table{my_bins_oneline_1.4}.  
The 
1994 observation appears to have caught the source in a low-intensity state.

In contrast, the 1997 mid-eclipse spectrum has relatively strong continuum emission.  The 2--10~keV 
flux is $1.1\times10^{-11}\rm\ erg\ cm^{-2}\ s^{-1}$.  Again, a model with two Gaussians was required to adequately fit the iron K-line region.  The results are shown in 
\Figure{my_bins_oneline_1.4_plot} and \Table{my_bins_oneline_1.4}.

\begin{table*}
 \centering
 \begin{minipage}{140mm}
  \caption{
Best-fitting model parameters for the mid-eclipse data of the 1994 and 1997 observations. 
The model is of the form shown in
\Equation{my_bins_oneline_1.4_equation} with two Gaussians
representing iron fluorescence and recombination emission lines. 
The data are for approximately the same orbital phases 
($-0.001\mbox{--}+0.074$ for the 1994 observation and
$-0.044\mbox{--}+0.011$ for the 1997 observation) and the same spectral binning has been applied.  Errors are for 90 per cent confidence in one interesting parameter.}
  \begin{tabular}{@{}ccccccc@{}}
  \hline
Parameter&&1994 Value&1997 Value\\
  \hline
SIS Exposure &(ks)&29.0&40.9\\
SIS count rate &(cts s$^{-1}$)&$0.142\pm0.002$&$0.026\pm0.001$\\
$N_{Hi}$&($10^{22}\rm\ cm^{-2}$)&$10.0\pm2.6$&$6.67 \pm 0.66$\\
$I_h$&($\rm\ cts\ cm^{-2}\ s^{-1}$ at 1 keV)&$(6.1\pm2.0)\times10^{-3}$&$0.028 \pm 0.012$\\
$N_{Hl}$&($10^{22}\rm\ cm^{-2}$)&$62\pm16$&$< 6.20$\\
$I_s$&($\rm\ cts\ cm^{-2}\ s^{-1}$ at 1 keV)&$(4.4\pm1.5)\times10^{-3}$&$(1.57 \pm 0.23)\times10^{-3}$\\
$\alpha$&&1.4 (fixed)&1.4 (fixed)\\
$E_{\rm fl}$&(keV)&$6.416\pm0.018$&$6.474 \pm 0.046$\\
$\sigma_{\rm fl}$&(eV)&$41\pm30$&$< 118$\\
$I_{\rm fl}$%
&($10^{-4}\rm\ cts\ cm^{-2}\ s^{-1}$)&$1.54\pm0.27$&$1.96 \pm 0.96$\\
$E_{\rm rc}$&(keV)&$7.13\pm0.15$&$ 6.83 \pm 0.17$\\
$\sigma_{\rm rc}$&(eV)&$260\pm190$&$< 380$\\
$I_{\rm rc}$%
&($10^{-4}\rm\ cts\ cm^{-2}\ s^{-1}$)&$0.92\pm0.39$&$2.3 \pm 1.1$\\
$\chi^2_\nu/\mbox{d.o.f}$&&$0.983/45$&$1.074/46$\\
\label{my_bins_oneline_1.4}
\end{tabular}
\end{minipage}
\end{table*}

\begin{captionfigure}{my_bins_oneline_1.4_plot}[]{
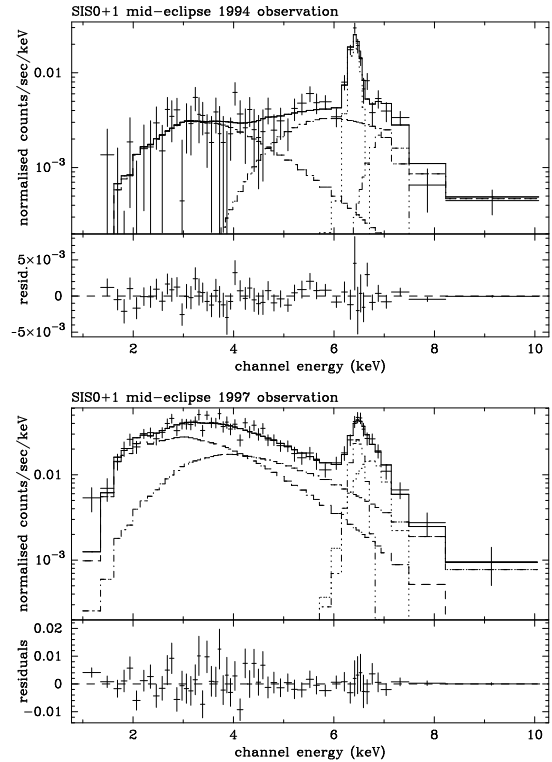
Top: SIS0+1 mid-eclipse spectrum and model for 1994 observation with
fit residuals.  Bottom: SIS0+1 mid-eclipse spectrum and model for 1997
observation with fit residuals.  The data are for approximately the
same orbital phases ($-0.001\mbox{--}+0.074$ for the 1994 observation and
$-0.044\mbox{--}+0.011$ for the 1997 observation) and the same spectral binning has
been applied.  The model parameters are as shown in
\Table{my_bins_oneline_1.4}.  Note the much stronger continuum in the
1997 observation.}
\plotfiddle{figure7.epsi}{284.760pt}{0}{57.3}{57.3}{-170.455pt}{-114.019pt}
\end{captionfigure}

\subsection{Spatial analysis}
From the slow decay of the soft-band lightcurves through the eclipse we suspected that an 
extended dust-scattered halo might be present.  In order to investigate this we extracted 
SIS and GIS images in various energy bands.  Surface brightness profiles were generated 
for the SIS and GIS data.  The SIS was operated in one-CCD mode which is not optimal for 
studying structures extended over more than a few arcmin.  This means that if we try to 
integrate a surface brightness profile over more than about 4~arcmin, we start including 
areas off the chip.  Combined with the considerable azimuthal and energy dependence of 
the XRT, this complicates spatial analysis.  Thus we used the {\sc ASCA-ANL} software to generate 
ray-traced simulated images which we could compare with the data.  In each case we simulated 
the image expected from a point source at the centroid coordinates of the observed source, 
and with the same spectrum.  We then integrated the data to produce a radial surface brightness 
profile (SBP).  We integrated the the ray-traced image in exactly the same way, from exactly the same region, to produce a 
SBP which represented the point spread function (PSF) expected from a point source at the 
same position with the same spectrum.  We parameterised the PSF with three exponentials and 
fitted for the widths and relative normalizations.  We then fitted the resulting 
three-exponential model, along with a constant background, to the SBP obtained from the 
data, keeping the widths and relative normalizations fixed.  The fits were performed only 
for the innermost and outermost one arcmin of the SBP, where the point source and the background should
dominate, respectively.

\Figure{twoprofiles} shows surface brightness profiles from SIS0 for the 0.5--3 and 5--10~keV 
energy bands fitted to the PSF and a constant background.
Comparing these, it is clear that the low-energy data require a spatially extended source, while 
the high-energy data are consistent with a point source and the XRT PSF.  We interpret this as 
evidence for a dust-scattered halo.

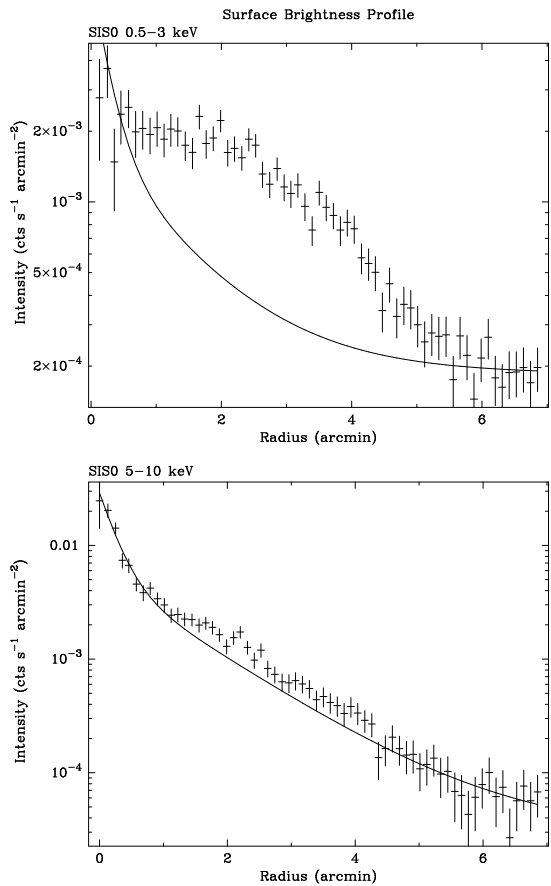
\begin{captionfigure}{twoprofiles}[]{SIS0 surface brightness profile
    for the mid-eclipse phase of the 1997 observation.  The data are fitted to the XRT PSF plus a constant for radii less than 
1~arcmin and between 6 and 7~arcmin.  Top: 0.5--3~keV.  Bottom:
5--10~keV. 
}
\plotfiddle{figure8.epsi}{330.033pt}{0}{65.6}{65.6}{-201.432pt}{-125.977pt}
\end{captionfigure}

\begin{figure}
\plotfiddle{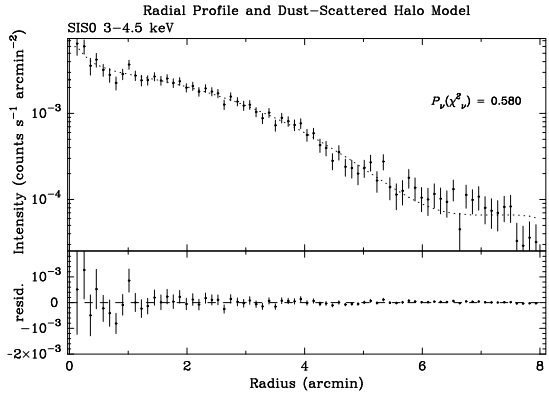}{145.541pt}{270}{29.8}{29.8}{-106.770pt}{167.313pt}

   \caption
   { \label{fig:halo}    
3.0--4.5~keV surface brightness profile for OAO 1657-415 for the
mid-eclipse phase of the 1997 observation.  The data have been fitted to a central point source, constant background, and dust-scattered halo.} 
\end{figure}

We can obtain a rough measure of the halo intensity from the residuals of the fits but for a 
more quantitative analysis we decided to fit the SBPs to model haloes, convolved with the 
ray-traced PSF.
We fitted the SBPs to a central point source, constant background, and dust-scattered halo.  
In the Rayleigh-Gans approximation the radial dependence of the model halo is $[j_1(x)/x]^2$, where $j_1(x)$ is the spherical 
Bessel function of order 1 and $x=(4\pi a/\lambda)\sin(\theta/2)$ \citep{Haya70}.  Here $a$ 
is the grain radius, $\lambda$ is the X-ray wavelength, and $\theta$
is the angle through which the dust grain scatters the X-ray photon.
The Rayleigh-Gans approximation is valid provided that $4\pi
a/\lambda\sin(\theta/2)\ll1$, which means in practice that we should
restrict our analysis to energies above 1~keV \citep[e.g.][]{Smith97}.  The observed angular distance of the detected X-ray photon from the centre of the image is $\alpha = (1-z)\theta$, where $z$ is the distance from the observer to the dust grain divided by the distance from the observer to the source.  The model was convolved with the ray-traced PSF for the fit.  To simplify computation, instead of using the Bessel-function form above we used the Gaussian approximation of \citet{Mauc86} for the halo shape.  
Figure~\ref{fig:halo} shows typical fit results for an SIS0 eclipse halo.  We divided the SIS0 and SIS1 eclipse data into energy bins ( 1.5--2.5, 2.5--3.5, 3.5--4.5, 4.5--5.5, and 5.5--7.0~keV) and obtained the best-fitting value of $4\pi a/\lambda$ for each energy bin, assuming that the dust grains were distributed uniformly along the line of sight.
We then plotted $4\pi a/\lambda$ against energy and the slope yielded the value
  $0.095\pm0.005\rm\ \mu m$ for the size of the dust grains.

The size distribution of the grains may be found either from the time dependence of the integrated halo spectrum \citep{XMcCK86} or from the surface brightness profile at a given energy \citep{Mauc86}.  Since we  find it easier to separate the central source from the halo in discrete energy bands by deconvolving the surface brightness profile from the point spread function, we adopt the latter approach.  We do not have good enough counting statistics for the former method.

\begin{figure}
\plotfiddle{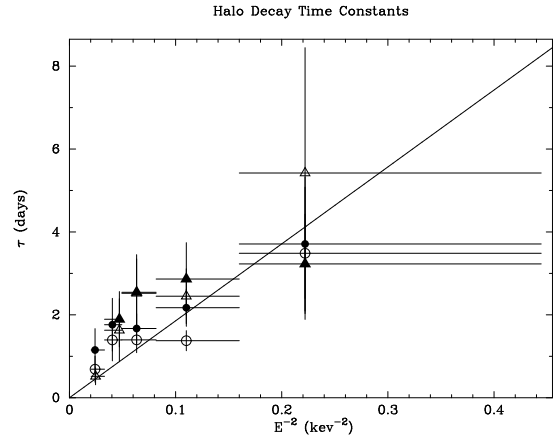}{161.878pt}{270}{30.1}{30.1}{-108.620pt}{173.612pt}
   \caption
   { \label{figure_decay}    
Halo decay time-constants plotted against the inverse square of the energy.  The SIS0, SIS1, GIS2, and GIS3 data are represented by open circles, filled circles, open triangles, and filled triangles, respectively.  The data have been fitted to a function of the form $t_d(E)=\tau E^{-2}$ and the best-fitting value is $\tau=18.6\pm2.3\ \mbox{d}$.} 
\end{figure}

We split the SIS0 and SIS1 eclipse data into 20~ks (gross) bins and extracted surface brightness profiles for different energy ranges ( 1.5--2.5, 2.5--3.5, 3.5--4.5, 4.5--5.5, and 5.5--7.0~keV).  We obtained the halo intensity for each time and energy bin.  We then fitted these halo intensities with an exponential function for each energy bin.
The spatial resolution of the GIS is such that the profiles are more
smeared out.  This means that the GIS values for the decay time
constant have larger error bars and in some cases are not constrained
by the data.  Thus, although the GIS results for the decay time
constant are consistent with the SIS results, we use only the SIS data
to obtain the decay time constant.  
If we assume the dust is uniformly distributed along the line of sight, we 
expect the halo to decay exponentially with a time constant of $t_d(E)=13.6E^{-2}D_8a_{0.1}^{-2}$~d, where $E$ is the photon energy in keV, $D_8$ is the distance to the source in units of 8~kpc, and $a_{0.1}$ is the dust 
grain radius in units of $0.1\rm\ \mu m$.  A fit of this form to the data is shown in \Figure{figure_decay}.  This results in a distance of 
$(7.9\pm0.9)a_{0.1}^2\ \mbox{kpc}$ for \oao.  We have implicitly
assumed here that the eclipse is a step-function.  The eclipse ingress
is much shorter than the eclipse duration and, as our measurements of
the source intensity start just before eclipse ingress, it seems
reasonable to assume a constant intensity prior to the eclipse
ingress.  However, this estimate of the distance is sensitive to the
distribution of the dust along the line of sight.  If the dust is all
in a sheet half-way to the source our estimate for the distance becomes
$(12.7\pm1.5)a_{0.1}^2\ \mbox{kpc}$, while it is
$(4.56\pm0.54)a_{0.1}^2\ \mbox{kpc}$ if the sheet is at either 10 or
90 per cent
of the distance to the source.

\section{Discussion}
\subsection{Intensity States}
\label{intensitystates}
We have observed \oao\ during the latter half of its eclipse in two
distinct intensity states and have found differences in the spectra.
The low-state spectrum is consistent with an absorbed power law with
photon index $0.36\pm0.19$ due to scattered and absorbed emission from
the pulsar, combined with a dust-scattered halo and fluorescent
emission from cold iron.  The high-state spectrum has a stronger
continuum which is due to scattering of the direct pulsar emission in
the stellar wind combined with the dust-scattered halo. 
In the high state the recombination line due to ionised iron is
stronger than the fluorescent line.  In the low state the fluorescent line is stronger. 
This difference between the two states is most likely due to increased
photoionization of the stellar wind in the high-intensity state.  

\subsection{The spectra}
The aim of the 1997 observation was to use the eclipse of \oao\ to study the outer 
atmosphere of the OB companion star and its vicinity by mapping out
the X-ray emission in the system. As the eclipse 
ingress progresses, the X-ray emission must propagate 
through the wind and outer atmosphere of the OB star. From the
progress of the partial
eclipses of the iron emission lines we conclude that the 6.4-keV fluorescent
emission originates near the neutron star while the 6.9-keV
recombination emission comes from the extended wind of the companion.

We find that a single power law cannot describe the continuum spectrum
outside of eclipse.
We obtained an acceptable fit using two power laws with the same photon index, 
one of which has a larger absorption column density than the other.  This 
model is equivalent to a partially-covered power law.  Similar two-component 
continua have been observed in Cen~X-3 \citep{Ken,myDiss}, Vela~X-1 
\citep{Sako1999}, and GX1+4 \citep{Endo2001}.   \citet{Sako1999} claim that the more absorbed component 
is the direct beam which is absorbed by the stellar wind surrounding the 
neutron star and that the less-absorbed component is due to 
scattering in the stellar wind.  \citet{Ken} claim the opposite: that the 
direct beam is the less-absorbed component.  One way to distinguish between 
these two pictures is by pulse-phase-resolved spectroscopy of the continuum.  
We would expect the component that is due to the direct beam to  pulsate at 
the pulsar frequency.  As long as the light travel time over the extent of 
the stellar wind is at least a significant fraction of the pulse period, 
pulsations in the scattered component should be smeared out.
\citet{Endo2001} did this in the case of GX1+4 and found that both
components originate near the neutron star.  \citet{myDiss} came to
the same conclusion for Cen~X-3.  When we fit the pulse-phase resolved
non-eclipse SIS data with the two-power law model, we find that the
intensities of the power laws have essentially the same dependence.
We therefore conclude that both of these components must originate at
the same place, significantly less than 38~lt-sec away from the neutron star.

\subsection{The dust-scattered halo}
Scattering of X-rays by interstellar dust leads to the formation of a halo 
whose intensity, shape, and spectral properties depend on the distance to the 
source, its intrinsic spectrum, the distribution of dust grains, and their 
size and composition \citep{Haya70}.  \citet{Trum73} proposed that the 
time-dependence of a dust-scattered halo from a variable X-ray source
could be used to determine the distance to the source and the
distribution of dust along the line of sight.  \citet{XMcCK86}
suggested that high-mass X-ray binaries with their abrupt total
eclipses that occupy a large fraction of the orbit would be good
candidates for this kind of study.  \citet{Mits90} used lunar
occultation observations with Ginga of two X-ray sources to study the
energy dependence of the dust halo and inferred the presence of iron
in the scattering grains.  \citet{Day91} applied the model of
\citet{Moln86} for the X-ray halo of a time-varying source to the low-energy eclipse
lightcurve of Cen~X-3 and obtained a distance of 5~kpc and a dust grain
radius of $0.33\rm\ \mu m$.  \citet{WC94} used the imaging capability of
{\sl ASCA's} SIS to measure the surface brightness profile of the
dust-scattered halo in Cen~X-3 and infer some properties of the dust
grains.  Applying the Rayleigh-Gans approximation, from the lower than
expected relative intensity of the halo they concluded that the dust
grains were loose aggregates rather than solid particles.  However,
\citet{Smith97} later pointed out that the Rayleigh-Gans approximation
for the differential scattering cross-section overestimates the halo
intensity below 1~keV and that when the more exact Mie solution is
used the data may be consistent with solid grains.  Later,
\citet{Drain03a} pointed out that the median scattering angle for the
dust halo would be small enough that most of the halo photons would be
seen in the core of the {\sl ASCA} image which is dominated by the central
point source.  This would have led \citet{WC94} to underestimate the
halo intensity and is likely to be a problem with the current
observation of \oao..
\citet{Pred00} demonstrated that Chandra could be used to determine
the distance to Cyg~X-3 to within 20 per cent by correlating the lightcurves
at different radii from the source. Cyg~X-3 only undergoes a partial
eclipse while high mass X-ray binaries such as \oao\ undergo total
eclipses for large fractions of their orbital periods.  This makes a
high-mass X-ray binary such as \oao\ a promising candidate for this
kind of study.  

We have found a decaying dust-scattered halo in the eclipse of \oao\
(see Figure~\ref{lightcurves}).  The half-power radius of the halo is
about 4~arcmin, which is not much larger than the XRT point spread
function that smears out the halo and mixes in flux from the core (see
Figure~\ref{fig:halo}).  The X-ray spectrum of \oao\ in eclipse has
three components: the dust halo, X-rays from the eclipsed pulsar
scattered by the primary's atmosphere, and emission (mostly iron-line
recombination) from the primary's atmosphere.  Only the dust halo will
appear to be extended in an X-ray image.    Because this source is heavily absorbed
($N_H\sim5\times10^{22}\rm\ cm^{-2}$) the halo is mostly seen above
the {\sl ROSAT} pass-band.  

\citet{Chak93} used the observed values of the pulse period and its
derivative during a steady spin-up interval to place a lower limit of
$\sim11\rm\ kpc$ on the distance to \oao.  \citet{Chak02} found a distance of
$6.4\pm1.5\rm\ kpc$ from the reddening of the optical counterpart.
The distance we estimated from the decay of the halo, $(7.9\pm0.9)a_{0.1}^2\ \mbox{kpc}$, lies between these two values.
 .
It is interesting to compare the case of Cen~X-3 where \citet{Day91} obtained a distance of $5.4\pm0.3\rm\ kpc$ which is a little lower than the lower limit derived from the relation between temperature and luminosity for the BO star \citep[6.2~kpc;][]{Krz}.
The value we obtained for the maximum dust-grain radius is $0.095\pm0.005\rm\ \mu m$ which is smaller than the value obtained by \citet{Day91} ($0.34\pm0.01\rm\ \mu m$).
There may be systematic effects resulting from our simplistic assumptions which causes us to underestimate the distance and overestimate the grain size.
We would like to point out, however, that the distance and grain size
estimates are independent; the measured halo intensity does not depend
on the assumed grain distribution.  Our estimate of the distance is
sensitive to the exact distribution of the dust along the line of
sight; we have assumed that it is uniformly distributed.  Our results also depend on the assumption that the intrinsic intensity and spectrum of the source were constant before eclipse.

\section{Conclusions}

The eclipse spectrum of the source in its high-intensity state is dominated by the decaying dust-scattered halo.  We estimate the radius
of the grains to be $0.095\pm0.005\rm\ \mu m$ and the distance to the
source to be $(7.9\pm0.9)a_{0.1}^2\ \mbox{kpc}$, where $a_{0.1}$ is
the grain radius in units of $0.1\rm\ \mu m$.  Using our estimate for
the grain size, the distance to the source becomes $7.1\pm1.3\rm\
kpc$.  We find that the non-eclipse
high-state continuum spectrum cannot be modeled by a single power law.  A
more-absorbed component due to scattering by material near the pulsar
is required.  The iron lines underwent partial eclipses during the 1997
observation.  We estimate that 80
per cent of the 6.4-keV fluorescent emission originated less than $19\pm12$~lt-sec
away from the pulsar while the 6.9-keV recombination line was mostly emitted
by the extended, ionised stellar wind.  During the 1994 observation,
the continuum emission was much lower and the iron emission during
eclipse was dominated by the 6.4-keV flourescent line, which suggests
that the source was in a low intensity state.

\section*{Acknowledgments}
MDA was supported by a 
fellowship from the Japan Society for the Promotion of Science.
This research has made use of data obtained through
the High Energy Astrophysics Science Archive Research Center Online Service,
provided by the NASA/Goddard Space Flight Center.  
We would like to thank Naomi Ota and Ryo Shibata for help with the ray-tracing software.  We are grateful to the referee, Peter Predehl, for his helpful comments.

\bibliographystyle{mn2e}
\bibliography{astrophysics}

\end{document}